\documentclass[10pt, journal, compsoc]{IEEEtran}
 
\usepackage{gensymb}
\usepackage{tabularx} 
\usepackage{lipsum}   
\usepackage{array} 
\usepackage{booktabs}   
\usepackage[hyphens]{url}

 \usepackage{enumerate}

\usepackage{multirow}

\ifCLASSINFOpdf
   \usepackage[pdftex]{graphicx}
  \else
     \usepackage[dvips]{graphicx}

 \usepackage{amssymb,amsmath}
 
\fi

\usepackage[caption]{subfig}
\ifCLASSOPTIONcompsoc
  \usepackage[nocompress]{cite}
\else
  \usepackage{cite}
\fi

\begin{document}

\title{BLE Beacons in the Smart City: Applications, Challenges, and Research Opportunities}

\author{
        Petros~Spachos,~\IEEEmembership{Senior Member,~IEEE,}
        and~Konstantinos~Plataniotis,~\IEEEmembership{Fellow,~IEEE}

}

\IEEEtitleabstractindextext{
\begin{abstract}
    
Internet of Things (IoT)  helps to have every individual interconnected with their surroundings and to interact with them through smart devices. In recent years, Bluetooth Low Energy (BLE) technology has become very popular in the smart infrastructures, the medical and retail industry, and many more, due to its availability in a plethora of wireless devices. BLE is widely used in IoT devices, such as smartphones, smartwatches, and BLE beacons. Beacons are small, low cost and low power, wireless transmitters that bring attention to their location by broadcasting a signal with a unique identifier at regular intervals.  BLE beacons are a promising solution for many smart city applications: from proximity marketing to indoor navigation. However, they do pose security and privacy challenges. This work discusses the characteristics of BLE beacons, the applications that can benefit from them, and the challenges they pose while try to identify research opportunities and future directions.     
\end{abstract}

}

\maketitle
     
\IEEEpeerreviewmaketitle
 
\IEEEraisesectionheading{\section*{Introduction}}
\IEEEPARstart{B}{luetooth} Low Energy (BLE) beacons, commonly known as beacons, are devices growing in popularity and research potential. Their deployment is projected to reach 400 million deployed devices globally by the year 2020~\cite{deployed}. They can help with the next step from IoT devices to smart and social objects that interact with the users~\cite{atzori, zanella}. Beacons broadcast signals at certain intervals and within their transmission range. An analogy of the beacon's operation is with the operation of a lighthouse, which  represents a known location that can be uniquely identified by its light. Every ship that sees the light, they know about the existence of the lighthouse. However, the lighthouse neither communicates with the ships nor does it know how many ships see its light. Similarly, a beacon is broadcasting a radio signal to advertise to BLE-enabled devices its presence in the area. It is not able to communicate with the devices nor to identify how many devices are receiving its signal.
 
\begin{figure}[t!]
\centering%
\includegraphics[width=\columnwidth]{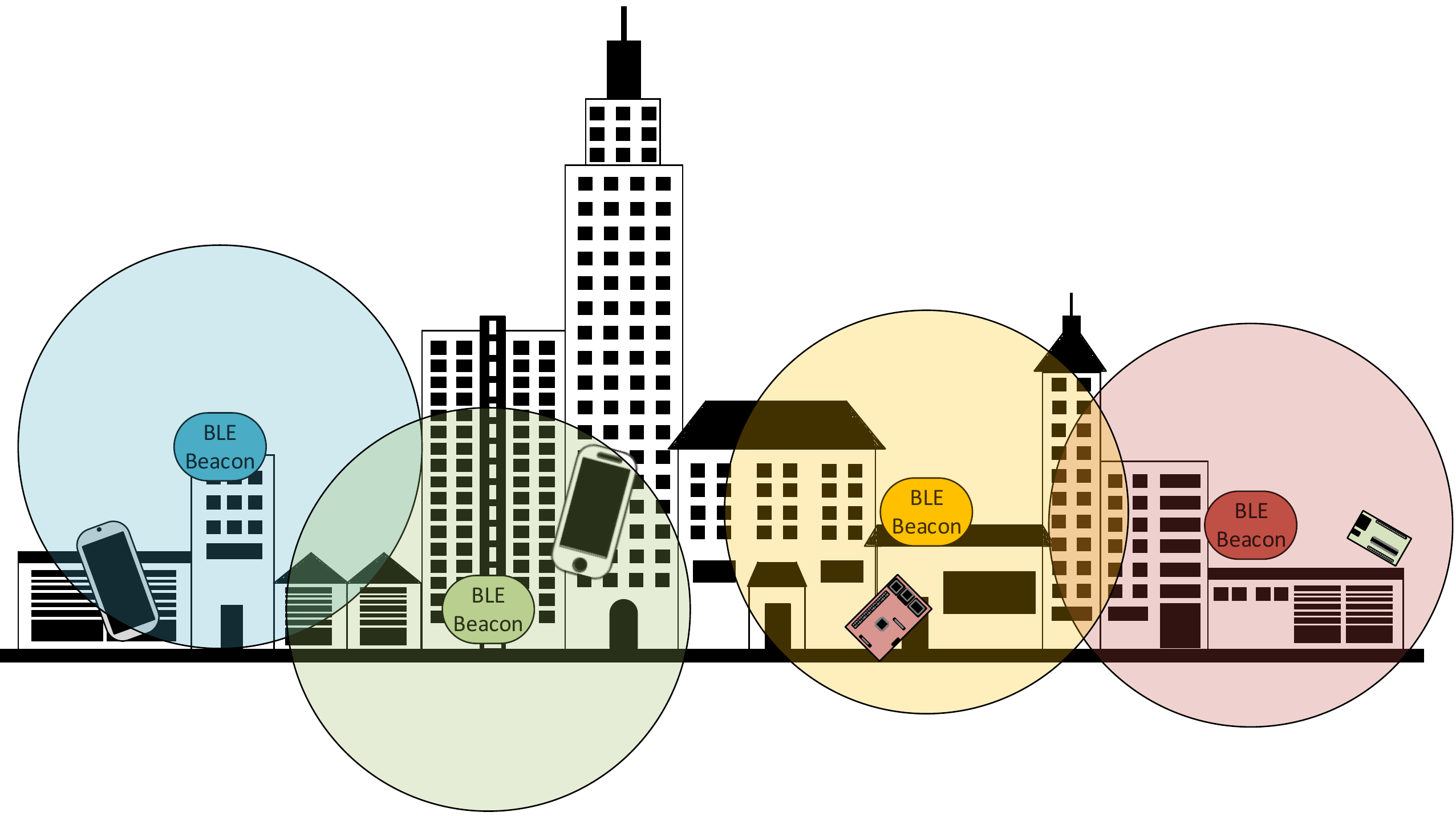}%
\caption{BLE beacons deployed in a city and broadcasting their signals to nearby devices. A device can listen signals from several beacons and take some actions in response.}%
\label{beac}%
\end{figure}

Beacons operation is shown in Fig.~\ref{beac}. Several beacons in an area, they broadcast their signals. BLE-enabled devices, such as smartphones, smartwatches and single-board computers like Raspberry Pis can listen to the signal and through applications, they can trigger some actions. These applications are running on the hosting device, while the beacons are not aware of them nor of the number of nearby beacon devices.

They are  wireless devices that use Bluetooth Low Energy (BLE) technology, which was developed for the purposes of low power consumption for applications that require minimal data throughput. Their small size, low cost,  and relatively long battery life increase their popularity. They broadcast packets that identify the particular beacon, along with possible telemetry data collected from sensors that may have been included by the beacon manufacturer. 

 Beacons are attractive solutions for several Internet of Things (IoT) applications~\cite{jeon}. Their small size and low cost provide a means of increased scalability and their BLE functionality establishes simple integration with smartphone devices, making them a highly versatile device.  They can be used in a plethora of smart city applications, from advertisement and Location Based Services (LBS) to indoor localization~\cite{spachos}, positioning, and tracking~\cite{he}.

\subsection*{Wireless Technologies for Smart City Applications}

For wireless data transmission, smart city applications and IoT devices  can take advantage of traditional wireless infrastructure, to minimize the additional deployment cost, or use some of the wireless technologies that are designed specifically for IoT devices and smart cities.  The unique characteristics of each application should always be considered before the selection of the proper technology~\cite{morin}.

Among the technologies that have been used successfully in the past for general purpose devices and they are popular for IoT systems as well, is the IEEE 802.11 standard, commonly known as Wi-Fi. A popular technology  due to the great distribution of access points and signal availability in different environments.  Zigbee is another popular communication protocol based on the IEEE 802.15.4 standard, known for its  low-power and secure networking.  Zigbee is intended to be simpler and less expensive than general wireless networking. LoRaWAN is a long-range, low power consumption wireless technology while Near Fields Communication (NFC) allows wireless data transfer between two portable devices in close proximity. Radio Frequency Identification Device (RFID) was primarily designed for data transferring and storing, and it can be passive (tags), where the electromagnetic field of the reader powers the device, or active (reader), where the RFID device has its own power source. Another popular technology is cellular IoT which connects IoT devices using existing cellular networks. Technologies such as NB-IoT and LTE-M will be a key part of 5G, which is a promising solution for future IoT applications with ultra-low latency and wide range services. There are also technologies that are specifically designed for IoT devices such as the IEEE 802.11ah (Wi-Fi HaLow) and the Bluetooth Low Energy (BLE) that were designed to support the concept of IoT and smart cities.

A popular wireless technology for short range communication is Bluetooth~\cite{bluetooth}. The standard is managed by the Bluetooth Special Interest Group (SIG) and can be found in several devices from mobile phones to robotic systems and laptops. Usually, it is used in symmetric connections between two devices. Bluetooth 4.0 aimed at novel applications in the healthcare, fitness, beacons and Bluetooth Low Energy (BLE) which is part of Bluetooth 4.0, is the popular beacon technology.  In comparison with traditional Wi-Fi,  it has low energy requirements and extended range. BLE was designed specifically for IoT and smart cities application~\cite{ble}. It has low power requirements and good data transfer rates.  BLE 4.0 can reach 25~Mbit/s at a distance of 60~meters. As a competitor of Wi-Fi HaLow among IoT devices,  Bluetooth 5.0 was introduced recently. This latest version is claimed to have four times longer transmission range, exchange data eight times faster, while it has twice the speed of the previous version.

\section*{BLE Beacons Characteristics}

BLE beacons have received a lot of attention due to their unique characteristics that made them ideal for several applications. They are wireless devices with the main goal of bringing attention to their location. Beacons are very small size, very low power, and especially low-cost devices that broadcast a wireless signal to all nearby devices. There is a wide variety of BLE beacon devices available from many vendors with different hardware, firmware, and protocol.

\subsection*{Hardware}
Beacon hardware is compact and simple. The hardware components dictate important factors, such as the cost, the power consumption, the performance, the on-chip memory, and the size. Similar to other wireless device hardware, beacon hardware has three components: the radio chip, the microcontroller,  and the power source. Additionally, there are beacons that have some sensors and general peripherals.

  \begin{table}[t!]   
\normalsize
    \centering
    \begin{tabular}{|l|c|c|c|c|}\hline
          \multirow{2}{*}{Manufacture} &  \multirow{2}{*}{SoC} & Integrated   & Current  \\
        & & Processor &Cons. (RX/TX)   \\ \hline  \hline
          \multirow{2}{*}{Texas} & CC2541 & 8051  & 18.2 mA \\ \cline{2-4}
            & CC256x & External & - \\ \cline{2-4}
                Instruments  & CC26xx & Cortex-M3 & 5.9 mA \\ \hline \hline

        Nordic & nRF51822 & Cortex-M0 &  9.7/ 8 \\ \cline{2-4}
     Semiconductors  & nRF8001 & External &  14.6/12.7 mA \\ \hline  \hline 
        Dialog  & \multirow{2}{*}{DA14580} &\multirow{2}{*}{Cortex-M0} &  \multirow{2}{*}{3.6 mA } \\ 
              Semiconductor && &   \\ \hline  \hline 
 
         Cypress &  \multirow{2}{*}{PRoC} & \multirow{2}{*}{Cortex-M0} &  \multirow{2}{*}{15.6/ 16.4 mA } \\  
             Semiconductor &  & &  \\   \hline 

      \end{tabular}
   \caption{Chipset and their characteristics.}
            \label{blechipset}

\end{table}

Texas Instruments, Nordic Semiconductors, Dialog Semiconductors, and Cypress are leading the beacon radio chip development process. Table~\ref{blechipset}, shows some representative popular BLE chipsets. The availability of an integrated processor, the flash and RAM capacity and the current consumption are important factors for the proper chip selection.  The processor in some of these chipsets is on-chip while some of them come without a processor. The 8051, and the ARM Cortex-M0 and Cortex-M3  are popular choices. For many smart city applications, this should  be sufficient, however, when more performance is required, standalone devices should be selected, that can work with an external microcontroller. The flash capacity starts from 32~kB and goes up to 256~kB and the RAM capacity is between 8~kB and 64~kB. It is important to note that all the chipsets support BLE v4.1 or v.4.2 which is the most common today. 

The power source is another critical internal component. Coin cell batteries are popular among beacons. Depending on the size of the hardware, the battery size varies. There are coin cell batteries of 240~mAh  allowing the beacon device to have very small dimensions, at the cost of reduced battery life. On the other hand,  standard AA~batteries of 2,000~mAh can be used to drastically improve battery life, at the cost of a far larger dimension. There are also  beacons with build-in Li-ion battery as well as solar-powered beacons. When external power is required, power outlet and USB outlet are popular choices.

\subsection*{Firmware}
Each beacon has a specific firmware that makes use of the available hardware. A critical characteristic of beacon-based applications is the lifespan of the beacon nodes. The firmware controls several characteristics that impact the total power consumption. Two are the main configuration parameters that greatly affect the power consumption: the transmission power, and the advertising interval. 

\begin{figure}[t!]
\centering
\includegraphics[width=\columnwidth]{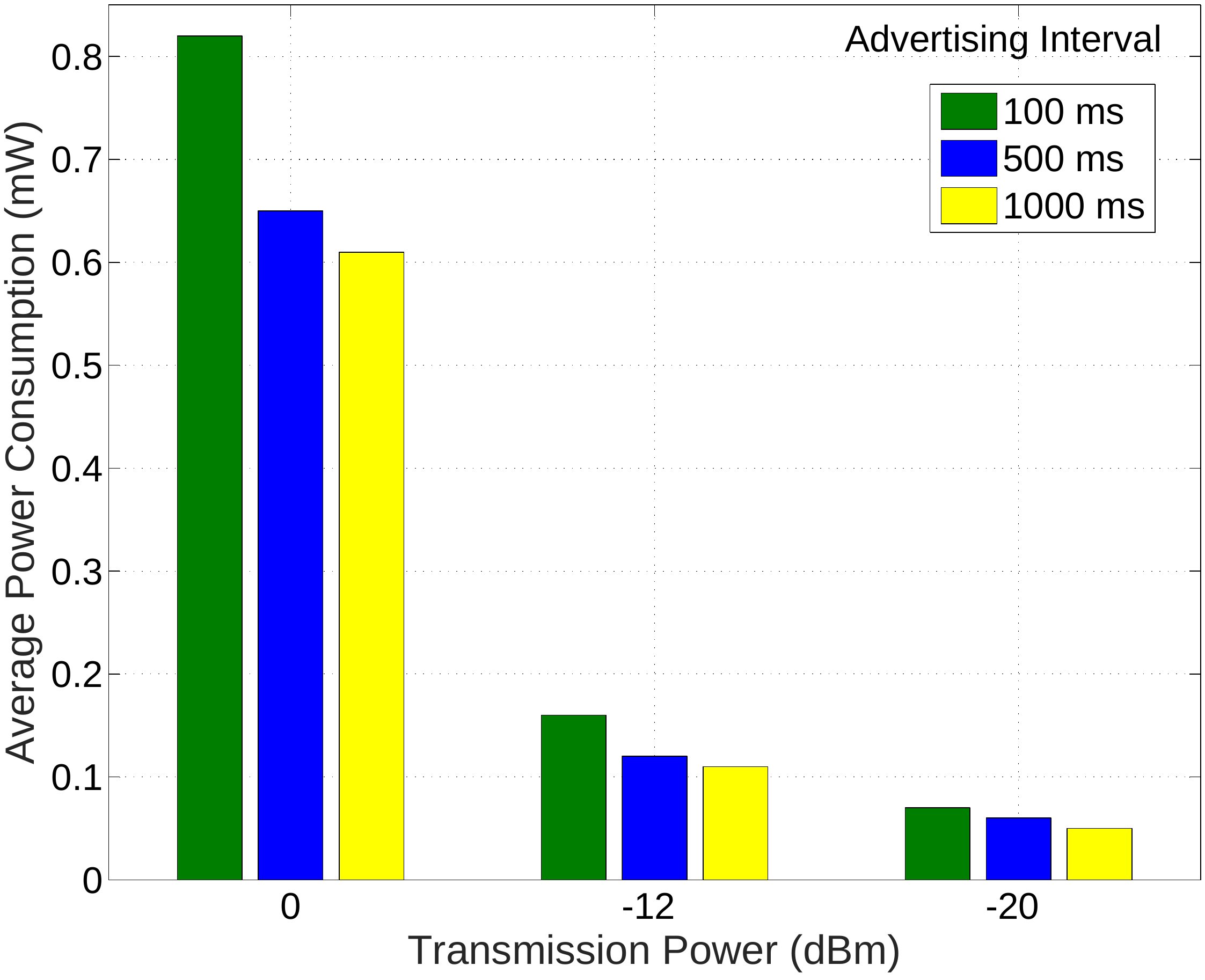}%
\caption{Average power consumption under different advertising intervals and transmission power levels.}%
\label{powerRelation}%
\end{figure}

Transmission power is the strength of the signal being broadcast, often represented as decibels with reference to 1~mW, i.e. dBm. As the signal travels in the air, its received signal strength decreases. This is a trade-off. High transmission power levels can achieve long distances with more power consumption requirements, while low transmission power achieves small ranges but less power is required. At the same time, long transmission range can increase the interference between beacons, while short ranges might not fit the application needs.

The second configuration parameter is the advertising interval. This is the frequency in which packet transmissions occur, often expressed in milliseconds. This is another trade-off. A high advertising interval of 100~ms (i.e. 10 times in a second) will lead to a faster battery drain, but the receiver can get more signals and perform tasks with high accuracy, such as micro localization~\cite{spachos}. On the other hand, a low advertising interval of 1,000~ms (i.e. 1 time in a second) will lead to an extended lifespan of the beacon, but it should be preferred in applications that can cope with this latency, such as in proximity-based applications.  Fig.~\ref{powerRelation} depicts how the energy consumption changes over three transmission power levels  and three advertising intervals  for a BLE beacon. It is clear that the power consumption is proportional to the transmission power and inversely proportional to the advertising interval.

\subsection*{Protocol structure}
BLE has 40 physical channels in the 2.4GHz ISM band, each separated by 2MHz. BLE defines two types of transmissions, advertising and data transmission. Out of the 40 channels, three channels, 37, 38 and 39 are used for advertising and the rest for data transmission~\cite{bluetooth}. The three channels were selected to avoid conflict with Wi-Fi traffic in the area. It is important to note that beacons are connectionless devices, hence no device pairing is required. BLE defines a packer format for transmission. This format has four components: preamble, access address, Protocol Data Unit (PDU) and Cyclic Redundancy Check (CRC). 
 
\begin{figure}[t!]
\centering
\includegraphics[width=\columnwidth]{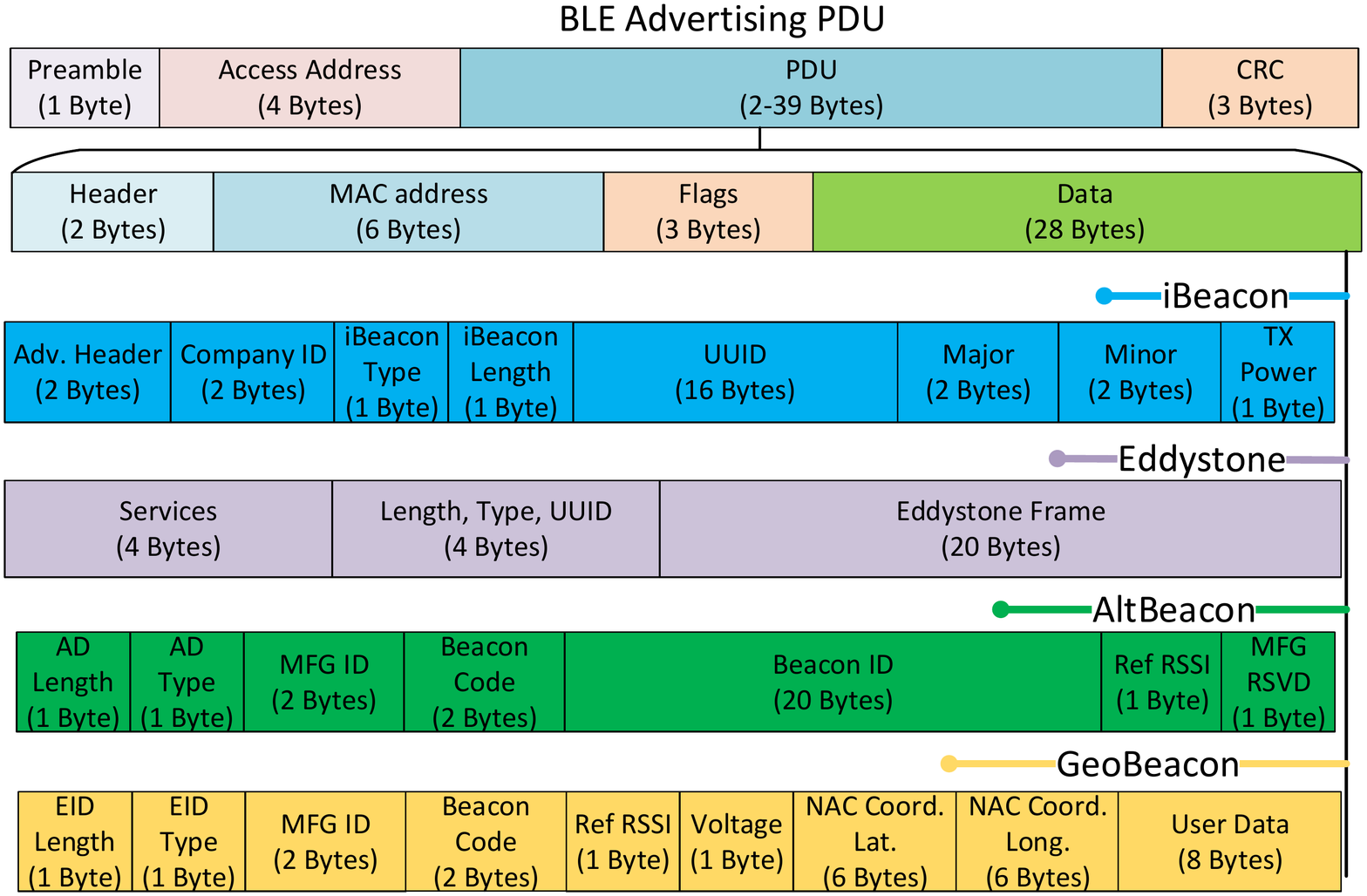}%
\caption{Structure of four BLE protocols.}%
\label{protocols}%
\end{figure}

Beacons need a protocol that facilitates the integration of manufacturing, programming, transmission, and general functionality. Bluetooth SIG has not defined an official beaconing standard, however, among the most commonly used protocols is the iBeacon by Apple~\cite{ibeaconpacket}, the Eddystone by Google~\cite{eddystone}, the AltBeacon by Radius Network~\cite{altbeacons}, and the GeoBeacon by Tecno-World~\cite{geobeacon}. The structure of each protocol is shown in Fig.~\ref{protocols}.  Among the different fields, the beacon ID is crucial, the Major and Minor fields can be used to identify different areas and applications while the RSSI and the coordinates provide useful information for positioning and tracking.

Each approach has advantages and for every smart city application, the protocol should be carefully selected. For instance, iBeacon and AltBeacon offer  more space to forward information. Although iBeacon uses specific manufacturer ID for every chipset, in AltBeacon, which is open source protocol, the manufacturer ID can be defined by the user. On the other hand, Eddystone broadcasts three different types of packets: UID, URL, and TLM which can help to transfer different data, while GeoBeacon has a very compact data storage and can provide high resolution coordinates, especially for location-based applications.

\subsection*{Additional Sensing Capabilities (Peripherals)}\label{sensing}
BLE beacons often include a variety of additional low power sensors such as temperature and humidity sensors, luxometer,  barometer,  accelerometer, gyroscope, microphone, etc. The sensors enable the beacons to provide useful information about the environmental conditions in which it is deployed. At the same time, they can be used for many unique applications, such as micro-climate data collection and acoustic monitoring level prediction.

\section*{Smart City Applications}
Several smart city applications can take advantage of the unique characteristics of beacons. They can be placed in many environments, both indoor and outdoor and convert them into smart areas, by providing interaction with the users. Most of the applications are context-aware services  and they fall into two general categories: Proximity-Based Services (PBS), and Location-Based
Services (LBS). In both services, beacons can be placed in a static position or attached to moving objects.

\begin{figure}[t!]  
\centering
\captionsetup[subfloat]{farskip=0pt}%
\subfloat[Beacon beside an exhibit.]{\includegraphics[width=0.45\columnwidth]{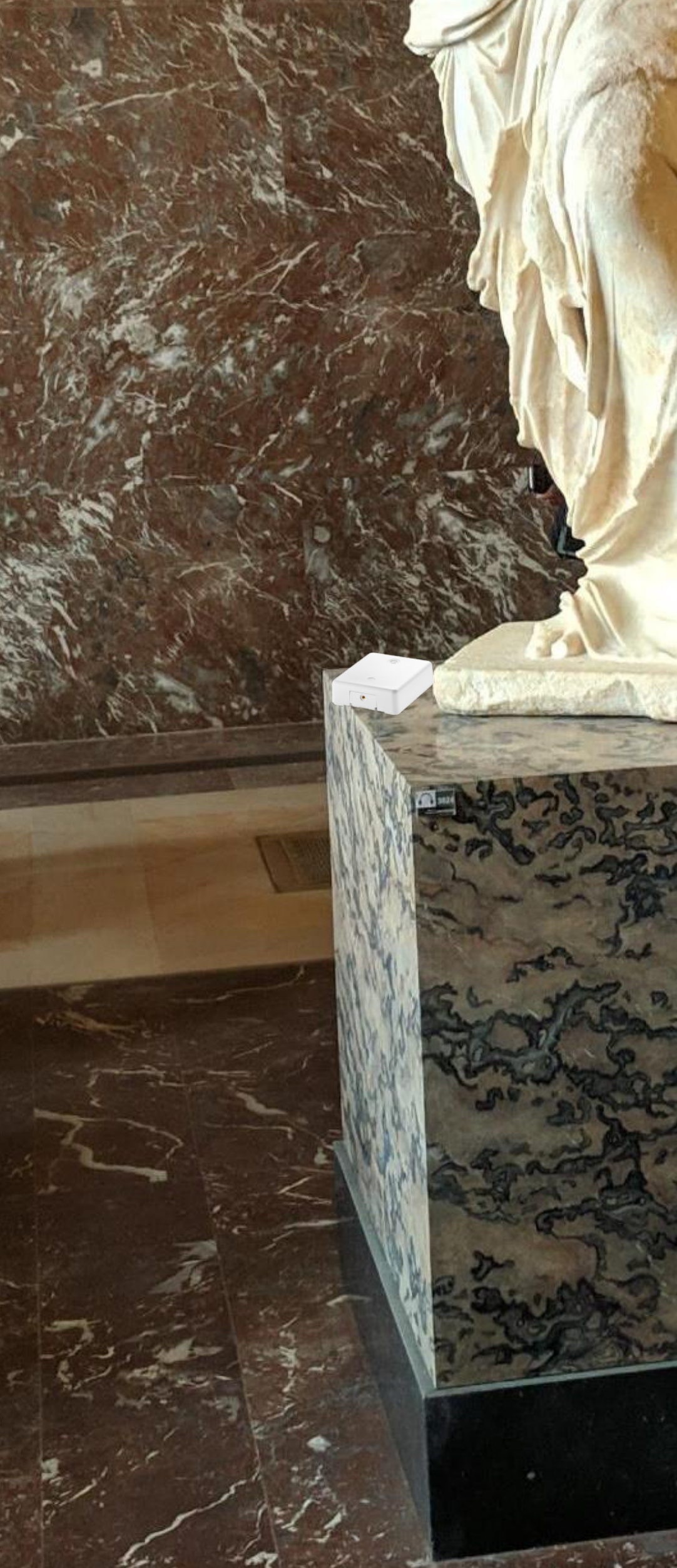}
\label{museum}}
\subfloat[Beacon inside a luggage.]{\includegraphics[width=0.45\columnwidth]{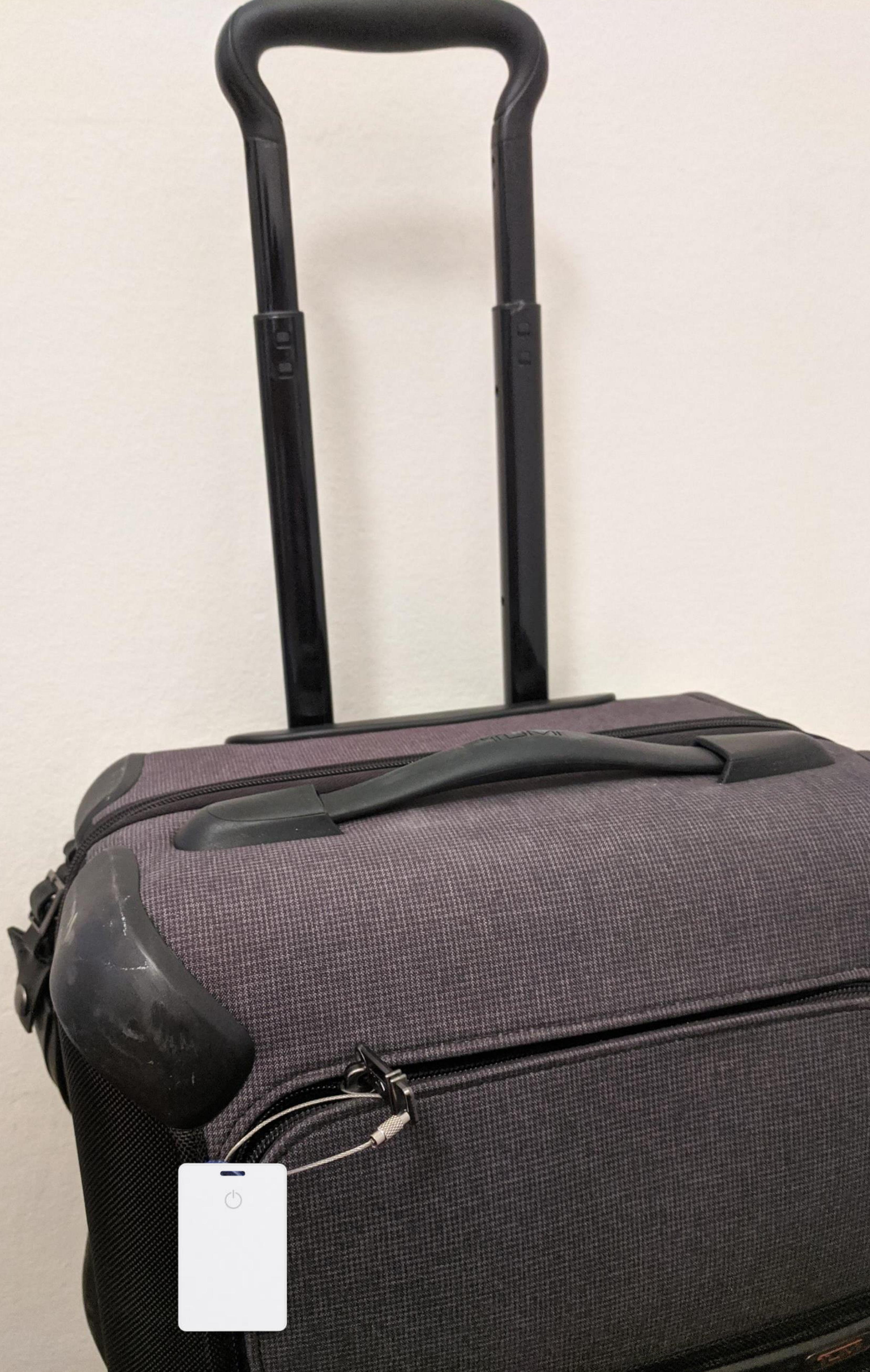}
\label{luggage}}
\caption{Proximity-Based Services: Beacon (a) besides Point of Interest and (b) attached to a moving object.}
\label{labraw}
\end{figure}

\subsection*{Proximity-Based Services}
PBS delivers information according to the proximity of the receiver node from the transmitter node. 

\subsubsection*{Point of Interest}
When the beacons are static, they can be used as Point of Interest (PoI) solutions and enhance interactivity in a smart museum or for proximity marketing. In a smart museum~\cite{museum}, they can be placed beside the exhibits, shown in Fig.~\ref{museum}, and when the visitors are close to them, they can forward to the visitor's smartphone, useful information about the exhibit. In a shopping mall, offers can be provided to the users when they are about to enter a store or a restaurant, as long as they have their smartphone Bluetooth active and use the proper mobile application. Beacons can offer many more opportunities to deliver context and enhance interactivity at the right time and place. However, the proper placement of the beacons and their advertising interval are crucial. The users might end up getting notifications  from every beacon in the area which can lead to too many notifications and information that might not be useful. 
 
\subsubsection*{Moving objects}
Beacons can be attached to moving objects such as luggage, bicycles or even cars. Mobile applications can be implemented in order to collect beacon signals and notify the users when they are close to them. For instance, beacons can be placed inside a luggage, shown in Fig.~\ref{luggage}, and when the luggage in a crowded airport is close to their owner, their smartphone can send them a notification. The advantages of using beacons in such applications are the low cost and the ease of deployment. At the same time, the broadcasting nature of beacon signals poses some security concerns. The location of valuable assets can be revealed to eavesdroppers through the beacon signals.

\begin{figure}[t!]  
\centering
\captionsetup[subfloat]{farskip=0pt}%
\subfloat[Indoor Positioning System.]{\includegraphics[width=0.5\columnwidth]{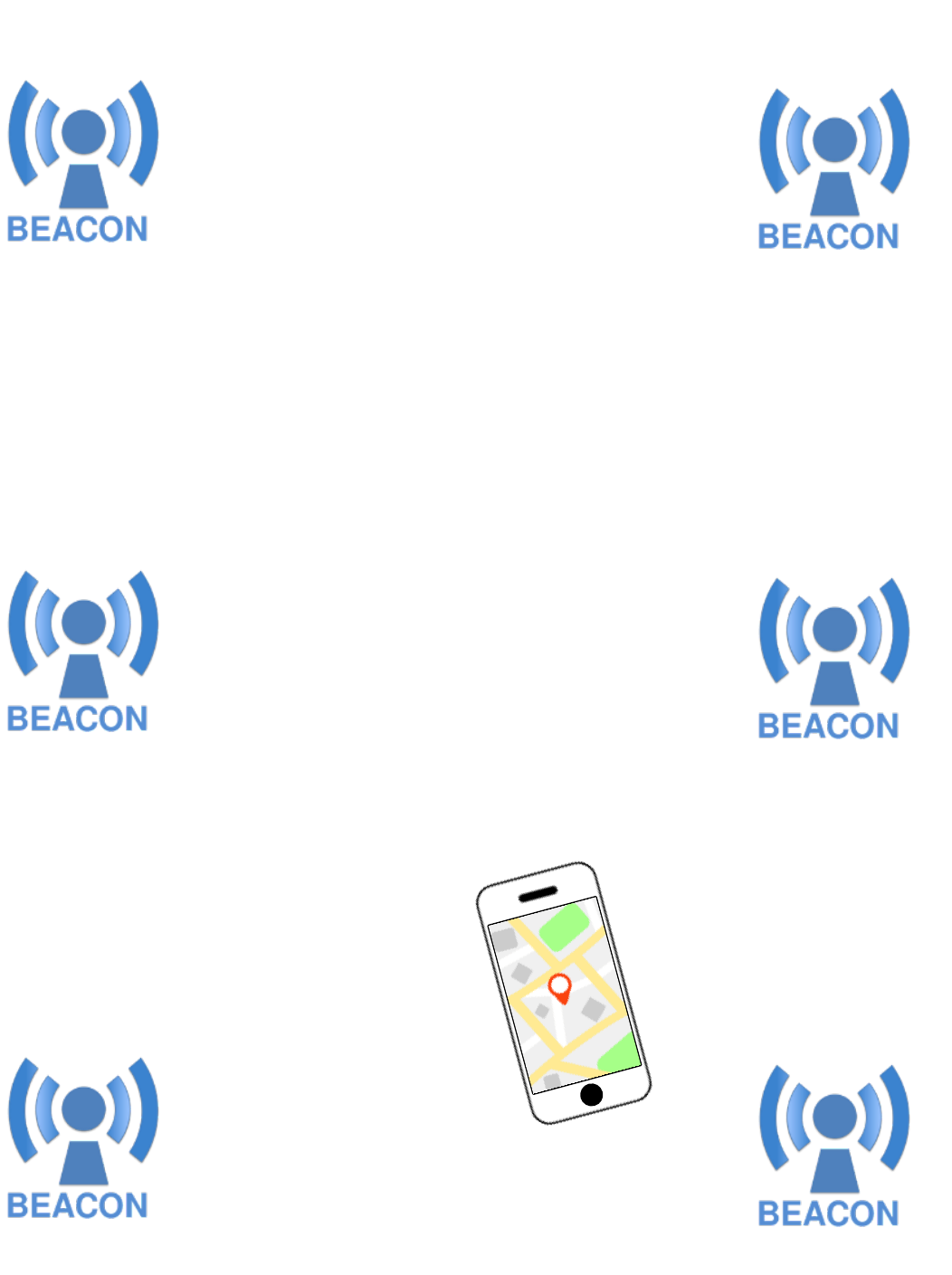}
\label{ips}}
\subfloat[Real Time Location System.]{\includegraphics[width=0.5\columnwidth]{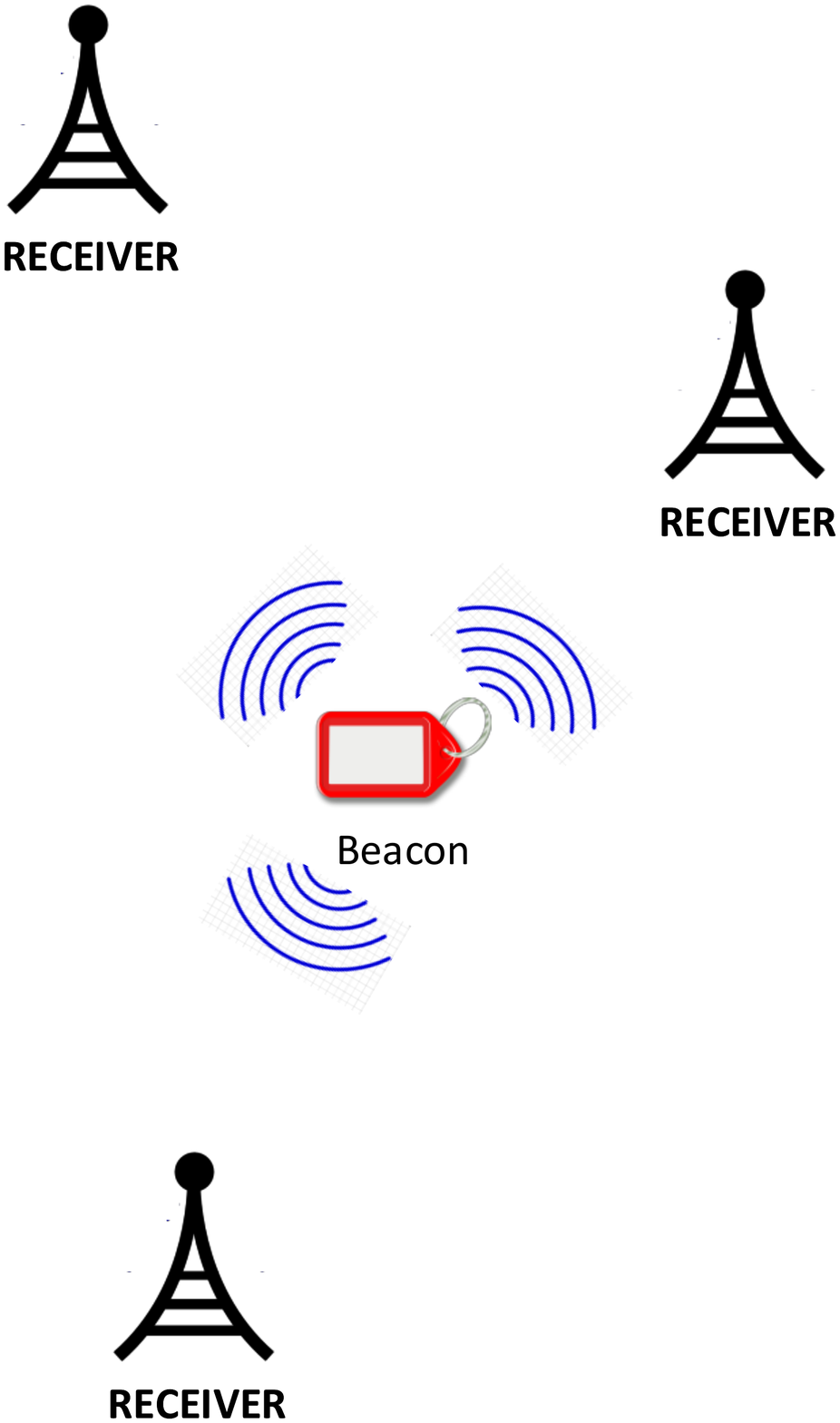}
\label{rtls}}
\caption{Location-Based Services: Beacon used (a) at an IPS and (b) at a RTLS.}
\label{labraw}
\end{figure}

\subsection*{Location-Based Services}
LBS delivers information according to the location of the user.

\subsubsection*{Indoor Positioning Systems}
In Indoor Positioning System (IPS), beacons can be deployed at static positions, shown in Fig.~\ref{ips}, in a complex indoor environment, such as an office or a University. A user can position in the area by performing localization on a BLE-enabled device such as a smartphone, according to the received signals. The beacons keep broadcasting while the users collect these signals with their smartphone and navigate in the area.   However, several factors, such as the number of the transmitting beacons and their transmission power can greatly affect the accuracy of the system. These factors should always be selected after extensive experimentation in the area.  The placement of the beacons would be a challenge, due to the fact the beacons performance decreases when the interference increases. Hence, beacons cannot be placed too close to each other. At the same time, techniques such as fingerprinting, which records the signal strength from several beacons in range and store this information in a database along with the known coordinates of each beacon, can be used to improve the overall system accuracy.

\subsubsection*{Real Time Locating Systems}
In Real Time Locating System (RTLS), beacons can be attached to valuable moving assets,  shown in Fig.~\ref{rtls}, in a hospital or a warehouse. RTLS works in the opposite way of the IPS: the moving beacon transmits the signal to edge devices, which performs the localization. A hospital can integrate beacon technology in order to maximize information exchange. Beacons can be placed on critical medical assets and devices. They can report the location of the assets in real time in a general management system. The beacon will transmit the signal to nearby collecting devices, such as Raspberry Pi.  Then, the beacon location can be estimated. There are many techniques that can be used to find the location of the beacon, such as trilateration from the Received Signal Strength of the beacons~\cite{spachos}. As the beacon is moving in the area, its location can be tracked based on the information it transmits to nearby devices. When needed, the exact location of each device in a large hospital can be found.  In such applications, the advertising interval of the beacon should be carefully selected to meet the expected moving speed of the object. At the same time, factors such as the number of the collecting devices and their placement can also affect the performance of the system. Advance filtering approaches should be implemented to improve the localization accuracy of the system. With further processing, beacons on nearby devices can also navigate the user to the required device.

\section*{Security and Privacy  Challenges}
    
The high deployment of beacons in smart city applications has also raised several security issues and privacy concerns~\cite{shao}. Some of these concerns are more challenges to cope with, depending on the application and the nature of the information.

\subsection*{Security issues}
There are some security issues regarding beacons. Some more challenging to cope with.

\begin{itemize}

\item \textbf{Cracking:} Due to their small size, beacons can be placed in many locations.  Usually, beacons are ``hidden" in different spots to cover an area. An important security attack on beacons is  their physical removal. An attacker can remove the beacon from a wall, open it up and have straight access to its hardware and any stored information. 

A real time monitoring of the beacon status can alleviate this problem  along with the decrease of the information that is stored inside the beacon memory. Information such as user passwords or user preferences should not be stored locally in the beacon. At the same time, when an interruption of the communication between the control system and the beacon happens, an alert should be sent to the system administrator. However, an increase in the communication between the beacon and the control system may affect the lifespan of the beacons dramatically. Therefore, there must be a balance between monitoring frequency and energy management.

\item \textbf{Spoofing:} Spoofing is when an attacker detects and clones a beacon. Beacons do not come with advanced encryption mechanisms. Hence, most of the time they broadcast their ID. An attacker who wants to attack the beacon and consequently, the users using this beacon, can use the same ID at another area and create a clone beacon. With the use of the clone beacon, the attacker can forward false information to the user. For instance, in a smart museum with beacons in the building, one beacon can send welcome messages to the visitors when they use the museum application and enter the building. An attacker can copy the beacon ID and replay the welcome message in  another location, far away from the museum, leading the visitors to remove the application.

A technique to minimize the issue of spoofing is by dynamically changing the ID of the beacons periodically. Beacons can create random IDs periodically to minimize spoofing. However,  the user has to accept a  connection to the new beacon ID every time.

\item \textbf{Piggybacking:}  Piggybacking is when an attacker listens to a beacon and captures the UUIDs, Majors, and Minors and adds them to another application without consent.  The attacker can then even clone the first application. For instance, in a shopping mall, Store A can offer a BLE-based mobile application that sends promotion codes to customers, when they are close. Store B can clone the beacon and the mobile application for its customers. In this way, when the customers that have the application of Store B, enters Store A, they will receive promotions for Store B.

\item \textbf{Hijacking:} In the communication with the beacons, there is no encryption techniques. Passwords or  important information that is broadcasted  from the beacon can be hijacked by an eavesdropper. Advanced encryption mechanisms can be applied to alleviate some of the security issues. However, this may adversely affect the lifespan of the beacon.

\end{itemize}

\subsection*{Privacy concerns}\label{privacy}
Privacy is important, especially when it comes to interaction with everyday objects and can reveal private patterns and habits. The following privacy concerns should be considered for BLE beacon applications.

\begin{itemize}
\item \textbf{Static IP.} Most BLE beacons have a static IP. This static IP is broadcasted so everyone in the transmission area can receive it. Hence, an attacker can mimic a trusted beacon, by using the same trusted IP and have access to private information. 

Many vendors have started research on dynamic IP assignment. This would require extra energy that may have varying effects on the lifespan of the beacon.

\item \textbf{Risk of unlawful surveillance.} Another important privacy concern comes from unlawful surveillance. Most of the beacon applications are based on localization. By using location services offered from beacons, the users share their location with them. 

Any attack on the beacons can reveal behavioural patterns about the user. Hence a user can be under surveillance without permission or their behavioural and location patterns can be shared with unauthorized personnel through the beacons.

\item \textbf{Risk of undesired advertisement.} Beacons are used a lot for advertisement. For more targeted advertisements, information about the user can be shared with the beacon applications. This information can be used by third parties for further undesired advertisements.
\end{itemize}

\section*{Future Directions}
Although some of the challenges remain, their unique characteristics will increase the BLE beacons deployment in smart city applications. Since beacons can transmit a lot information over seconds, providing in this way a large amount of data, machine learning techniques can be used to improve their usage. It is important to have smart data processing for smart city applications. For instance, deep learning approaches can be applied to improve the localization accuracy of the deployed systems. The availability  and the low cost of the BLE signals can help towards this direction. At the same time, general machine learning approaches can be applied to alleviate some of the current security and privacy challenged. The integration of machine learning approaches that can enhance security and privacy, into  content aware location-based services along can open a new and promising research area. Similarly, social learning using BLE beacons can be used to promote wellness.

The new BLE v.5.0 can improve the accuracy of current navigation applications and lead to the development of several more. The addition of the direction finding capability can be used along with techniques such as Angle of Arrival (AoA) and Angle of Departure (AoD) to improve the performance of the BLE-based systems. 

\section*{Conclusion}
 BLE beacons are a promising, low cost and energy-efficient IoT solution, mainly for location application. The selection of the proper beacon and the optimal configuration of the beacon parameters are important factors for successful application deployment. Experiments should be conducted to examine the performance of different beacons in different areas while security and privacy should always be a concern. At the same time, the unique characteristics of BLE beacon make them an attractive solution for several smart city applications.

\bibliographystyle{IEEEtran}
\bibliography{IEEEabrv,iotmagbib}

\begin{IEEEbiography}
[{\includegraphics[width=1in,height=1.25in,clip,keepaspectratio]{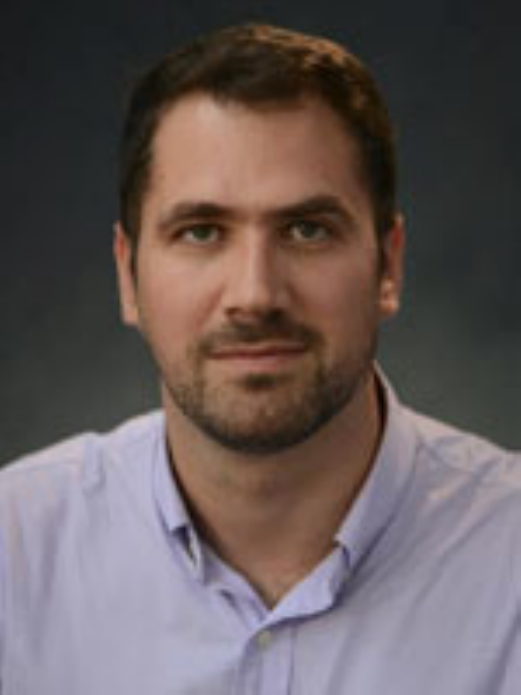}}]{Petros Spachos} (M'14--SM'18) received the Diploma  degree in Electronic and Computer Engineering from the Technical University of Crete, Greece, in 2008, and the M.A.Sc. degree in 2010 and the Ph.D. degree in 2014, both in Electrical and Computer Engineering from the University of Toronto, Canada.  He was a post-doctoral researcher at University of Toronto from September 2014 to July 2015. He is currently an Assistant Professor in the School of Engineering, University of Guelph, Canada. His research interests include experimental wireless networking and mobile computing with a focus on wireless sensor networks, smart cities, and the Internet of Things. He is a Senior Member of the IEEE.
\end{IEEEbiography}

\begin{IEEEbiography}[{\includegraphics[width=1in,height=1.25in,clip,keepaspectratio]{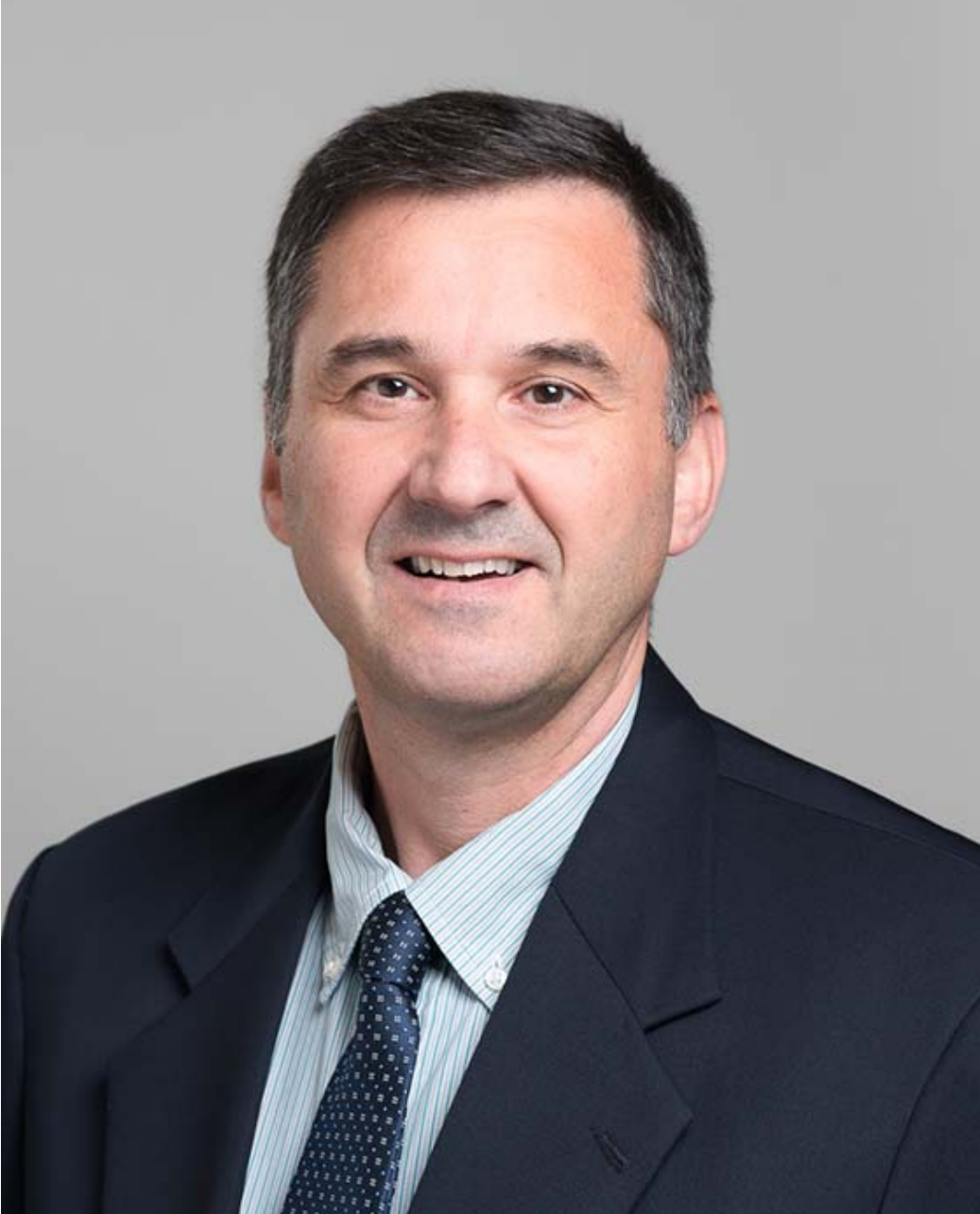}}]{Konstantinos N. (Kostas) Plataniotis}  (S'90--M'92--SM'03--F'12) received his B. Eng. degree in Computer Engineering from University of Patras, Greece and his M.S. and Ph.D. degrees in Electrical Engineering from Florida Institute of Technology Melbourne, Florida. Dr. Plataniotis is currently a Professor with The Edward S. Rogers Sr. Department of Electrical and Computer Engineering at the University of Toronto in Toronto, Ontario, Canada, where he directs the Multimedia Laboratory. He holds the Bell Canada Endowed Chair in Multimedia since 2014. His research interests are primarily in the areas of image/signal processing, machine learning and adaptive learning systems, visual data analysis, multimedia and knowledge media, and affective computing. Dr. Plataniotis is a Fellow of IEEE, Fellow of the Engineering Institute of Canada, and registered professional engineer in Ontario.

Dr. Plataniotis has served as the Editor-in-Chief of the IEEE Signal Processing Letters. He was the Technical Co-Chair of the IEEE 2013 International Conference in Acoustics, Speech and Signal Processing, and he served as the inaugural IEEE Signal Processing Society Vice President for Membership (2014 -2016) and General Co-Chair for the 2017 IEEE GLOBALSIP. He serves as the 2018 IEEE International Conference on Image Processing (ICIP 2018) and the 2021 IEEE International Conference on Acoustics, Speech and Signal Processing (ICASSP 2021) General Co-Chair.
\end{IEEEbiography}

\end{document}